\documentclass[twocolumn,showpacs,amsmath,amssymb,prb]{revtex4}
\usepackage{graphicx}% Include figure files
\usepackage{amssymb}
\usepackage{amsmath}

\begin{document}
\preprint{MR-28}
\title{Upper critical field from normal state fluctuations in Bi$_2$Sr$_2$CuO$_{6+\delta}$}

\author{F. Bouquet}
\author{L. Fruchter}
\author{I. Sfar}
\altaffiliation[Also at ]{L.P.M.C., D\'{e}partement de Physique,
Facult\'{e} des Sciences de Tunis, campus universitaire 1060
Tunis, Tunisia.}
\author{Z.Z. Li}
\author{H. Raffy}
\affiliation{Laboratoire de Physique des Solides, C.N.R.S.
Universit\'{e} Paris-Sud, 91405 Orsay cedex, France}

\date{Received: date / Revised version: date}
\begin{abstract}The in-plane magnetoresistance of an epitaxial Bi$_2$Sr$_2$CuO$_{6+\delta}$ thin film was systematically investigated as a function of doping, above $T_c$. The orbital magnetoconductance is used to extract the crossover field line $H_{c2}^*(T)$ in the fluctuation regime. This field is found in good agreement with the upper critical field obtained from resistivity data below $T_c$, and exhibits a similar upward curvature, thus pointing toward the existence of a critical correlation length. The consequences regarding the nature of the resistive transition are discussed.
\end{abstract}

 % end of PACS codes

\maketitle
\section{Introduction}
The superconducting transition in a magnetic field of high-$T_c$ cuprates has shown unusual features. Within a flux line description, the combined effects of high temperature, strong anisotropy, short coherence length and large penetration depth promotes a large flux liquid domain in the vicinity of the upper critical-field line, $H_{c2}(T)$. Such a liquid is characterized by the loss of the phase coherence, which is otherwise present in the ordered Abrikosov vortex lattice;\cite{nguyen1998,nguyen1999} the first order flux melting transition (or the irreversibility line, in the case of disordered materials) replaces the usual superconducting transition line, which is no longer characterized by the occurrence of zero resistivity. 

The absence of a genuine transition opens the possibility to describe the `normal' state as a region of fluctuating vortices. However, in the absence of local pairing, regular vortices do not survive above $T_c$. The observation of a large Nernst effect well above $T_c$, combined with a diamagnetic magnetization, has lent support to the idea that only the phase  coherence is broken at the superconducting transition, while the condensate amplitude remains finite.\cite{emery1995,xu2000} In the vortex description, this means that the vortex-core energy remains essentially unchanged at the superconducting transition and that it is unusually small. Therefore, the usual mirror in the descriptions of the fluctuating superconducting state and of the fluctuating normal state is broken: this results from the observation of an upper critical field which remains constant through $T_c$.\cite{wang2003,wang2005}

Within this picture, the superconducting transition line obtained from the saturation of the resistivity to a value close to the normal state one marks the loss of phase coherence. It roughly coincides with the peak in the Nernst effect (`ridge line') and it is strikingly different from the one inferred from the vanishing of the Nernst effect related the order parameter amplitude.\cite{wang2003,wang2005} Such a picture does not, however, completely clarify the superconducting phase diagram. In particular, there is in the case of Bi-based cuprates a large temperature interval between the onset for the Nernst signal and the temperature for its maximum and one may wonder whether a thermodynamic transition line for the loss of the phase stiffness exists, or whether the latter is gradually lost in this temperature interval.

Most of the speculations concerning the nature of the resistivity line were brought by the early observation of its pronounced upwards curvature for several cuprates and organic materials (see Ref.~\onlinecite{mackenzie1993} and refs therein). Experimentally, the reality of this curvature in the $H_{c2}(T)$ line is discussed. It was soon pointed out that this might result from an incorrect determination of the upper critical field from resistive measurements.  The in-plane resistive transition might yield a considerably lower value than the true upper critical field, due to the existence of a large flux liquid regime above the melting line:  zero-resistivity measurement would actually detect the irreversibility line (with an upward curvature)  instead of $H_{c2}$.  Indeed, in Refs.~\onlinecite{vedeneev1999} and~\onlinecite{vedeneev2006}, it was found that, for strongly overdoped and slightly underdoped Bi-2201 single crystals, a linear $H_{c2}(T)$ line is obtained, provided that the resistive criterion is chosen close to the normal state resistivity value. Existence of similar thermally activated motion of pancake vortices may be opposed to the critical field measurements from out-of-plane resistivity.\cite{koshelev1996} However, analysis of magnetization measurements on Bi-2201 showed that even though a linear temperature dependence of the upper critical field allows for the scaling of the magnetic moment with field and temperature,  a power law $(T_c-T)^{2.5}$ is needed to scale both the magnetic moment and its second derivative, $\partial^2M/\partial T^2$, thus implying a positive curvature in the $H_{c2}$ line.\cite{janod1996} 

Considering theories, some of them claim that the curved line is a genuine transition intrinsically related to the superconducting mechanism in cuprates. In Ref.~\onlinecite{alexandrov1996}, the curved $H_{c2}(T)$ obtained from out-of-plane resistivity is found consistent with the predictions based on the Bose-Einstein condensation of bosons formed above $T_c$. It was argued that this model is unable to account for the observation of similar data for overdoped samples, whereas this could be explained within the Boson-Fermion model, which accounts for both real-space paired carriers and itinerant Fermions.\cite{domanski2003} In Refs.~\onlinecite{ovchinnikov1996,ovchinnikov1999}, a peculiar magnetic pair-breaking mechanism specific to two-dimensional superconductors, allowing both strong pair-breaking effects and a clean situation, was shown to correctly describe the $H_{c2}(T)$ features. Finally, it was proposed that the critical field obtained from resistivity is actually the one for phase ordering of superconducting grains embedded in the material (with a critical temperature higher than the zero-field resistivity one);\cite{spivak1995,geshkenbein1998} above this field, decoupled grains with non-zero superconducting order parameter would subsist. This mechanism could be either intrinsic to these cuprates for which there is a local doping effect,\cite{davis2005,mashima2006} or could originate from chemical inhomogeneities or substitutions due to the elaboration process.

In this situation, experiments which are able to probe the upper critical field above and below $T_c$ are valuable. In the following, we demonstrate that magnetoresistance measurements reveal a crossover field line $H_{c2}^*(T)$ \textit{above} $T_c$ which is essentially similar to the $H_{c2}(T)$ line \textit{below} $T_c$. This suggests that both lines are governed by a common correlation length, as expected for a regular continuous transition.

\section{Experiments}
\subsection{Sample and experimental setup}

\begin{figure}
\includegraphics[width= \columnwidth]{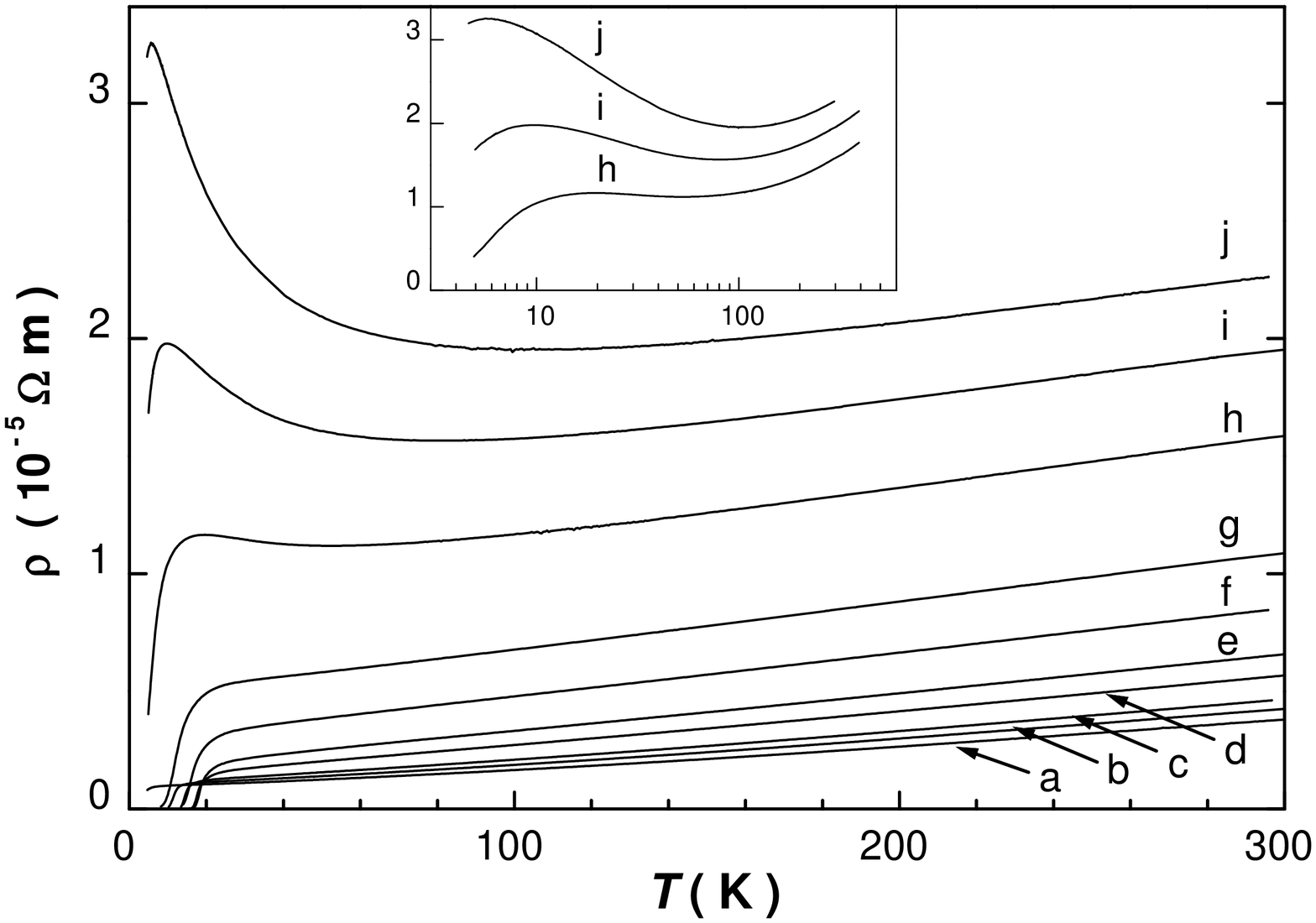}
\caption{$\rho(T)$ in zero field for different doping. The sample was fully overdoped; then successive doping states were obtained on the same sample by decreasing its oxygen content (see text for details). These states were labeled chronologically, from \textbf{a} the most overdoped, to \textbf{j} the most underdoped doping state. The optimally doped state corresponds to label \textbf{e}. Inset, the three most underdoped states are represented, in a semi-log representation.
}\label{RMR}
\end{figure}

\begin{figure}
\includegraphics[width= \columnwidth]{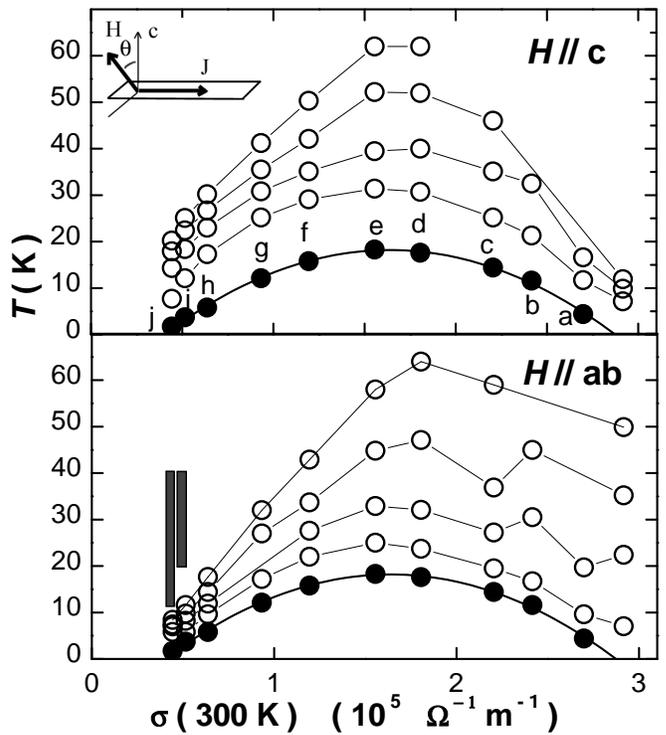}
\caption{Raw magnetoconductance data $\Delta\sigma(H = 6$~T$)$ for field applied perpendicular and parallel to the plane direction. The measurement geometry is displayed in the upper panel. Full circles: $T_c$ (the line is a parabolic fit). Open circles: the temperature where the magnetoconductance reaches a given criterion ($\Delta\,\sigma (6~T)=-9.5 \times 10^{-3}, -1.9\times 10^{-2}, -4.7\times 10^{-2}$, and $-0.19$ in units of $\sigma_0 = e^2/16 \hbar s$ from top to bottom. Shaded: area where a positive magnetoconductance is observed.}\label{MR}
\end{figure}

Transport measurements were performed on one epitaxial, 2700~\AA{} thick, $c$-axis oriented, Bi$_2$Sr$_2$CuO$_{6+\delta}$ (Bi-2201) thin film. Different levels of doping for this sample were studied, from superconducting overdoped to underdoped states (Fig.~\ref{RMR}). The sample was grown on a heated SrTiO$_3$ substrate, using \textit{rf} sputtering (Ref.~\onlinecite{li93} and refs. therein) and characterized by X-ray diffraction. The quality of the $c$-axis orientation is attested by the rocking curve through the (008) peak, which has a full width at half maximum less than 0.2$^\circ$. It was patterned in the standard 4 point resistivity bridge, with a 160~$\mu$m long and 95~$\mu$m large strip. The sample was first annealed at 420$^\circ$C and slowly cooled in a pure oxygen flow, yielding a non superconducting overdoped state. Successive lower-doped states were obtained by annealing the sample under vacuum, using temperatures between 240 and 290$^\circ$C, thus decreasing its oxygen content. Figure~\ref{RMR} presents the resistivity curves of these different doping states, labeled from \textbf{a} (most overdoped) to \textbf{j} (most underdoped).
Transition temperatures are defined at the midpoint of the resistive transition. The transition width (10\% -- 90\% completion) of the optimally doped state ($T_c \simeq 19$~K) was 2~K. The evolution of doping was followed through the evolution of these transition temperatures. We observe that the transition temperature as a function of the conductivity at 300~K is well described by a parabolic law (see Fig.~\ref{MR}). Although such a behavior is compatible with the empirical law relating the transition temperature to the doping level\cite{presland1991} and a linear relationship between the conductivity and the doping level, the conductivity should not be considered here as a direct measurement for the hole content as pointed out by Ref.~\onlinecite{ando2000}, but merely as a monotoneous function of it.

The resistivity was measured using a standard lock-in detection, with a 90~$\mu$A \textit{ac} current. Special care was taken to limit temperature shifts during measurements under a magnetic field: after stabilization of the temperature in zero magnetic field using a calibrated thermometer, the temperature measurement and regulation was switched to a capacitance sensor and the magnetic field was ramped from 0 to 6~T, with the field parallel or perpendicular to the planes. The temperature stability was better than 30~mK. Angular measurements -- varying the angle $\theta$ between the normal to the film and the applied field -- were performed to evaluate the sample anisotropy. The configuration of the measurement is displayed in the inset of Fig.~\ref{MR}. Using the minimum in the magnetoresistance, we were able to align the magnetic field within the film plane (CuO$_2$ plane) to better than 1$^\circ$ to determine the longitudinal magnetoresistance (with $\theta = \pi/2$) which, together with the transverse magnetoresistance (with $\theta = 0$), was used to obtain the orbital and the isotropic magnetoconductivity.

\subsection{Determination of the magnetoconductance components}

The raw data of the magnetoconductance $\Delta\sigma$ at 6~T are presented in Fig.~\ref{MR}, for $\theta=0$ and $\theta=\pi/2$. We found that a simple anisotropic-mass law\cite{blatter1992} alone cannot describe our data.
We postulate that the measured magnetoresistance originates from the sum of an isotropic magnetoconductance (independent of the magnetic field orientation), $\Delta\sigma_{iso}$, and an orbital one which scales with field and angle according to the anisotropic-mass law,\cite{blatter1992,larkin2004} $\Delta\sigma_{orb}$:
\begin{equation}
\Delta\sigma(\theta,H) = \sigma(\theta,H)-\sigma(\theta,0)=\Delta\sigma_\text{orb}(\tilde{H}) + \Delta\sigma_\text{iso}(H),
\label{scale}
\end{equation}
where $\tilde{H}=H\;[\cos(\theta)^2+\gamma^2\sin(\theta)^2]^{1/2}$ and $\gamma^2 < 1$ is the anisotropic mass ratio. So we have :
\begin{eqnarray}
\Delta\sigma_\text{orb}(H) &=& \Delta\sigma(0,H)-\Delta\sigma(\pi/2,H)+\Delta\sigma_\text{orb}(\gamma\;H)
\label{scalea}\\
\Delta\sigma_\text{iso}(H) &=& \Delta\sigma(\pi/2,H)-\Delta\sigma(0,\gamma\;H)+\Delta\sigma_\text{iso}(\gamma\;H)
\label{scaleb}
\end{eqnarray}

\begin{figure}
\includegraphics[width= \columnwidth]{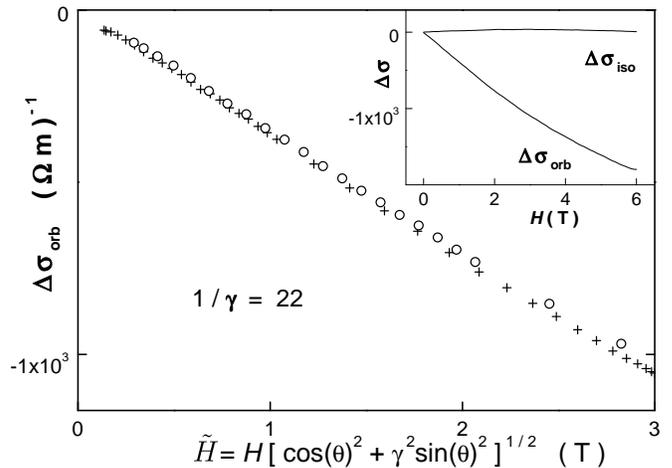}
\caption{Scaling of the orbital magnetoconductance ($T = 9$~K, $H$ = 3~T (crosses) and $H$ = 6~T (circles), varying $\theta$ from 0 to $\pi/2$) for the most underdoped state (\textbf{j} in Fig.~\ref{MR}), using the anisotropic mass law and $\gamma^{-1} = 22$. The inset shows the isotropic and orbital magnetoresistance, as determined by the iterative procedure.}\label{angular}
\end{figure}

As is usually done, we have neglected the last term in Eqs.~(\ref{scalea}, \ref{scaleb}) to obtain $\Delta\sigma_\text{orb}$ (Fig.~\ref{MR}, $H$//~ab) and $\Delta\sigma_\text{iso}$ (Fig.~\ref{MR}, $H$//~c) from the longitudinal magnetoconductance, $\Delta\sigma(\pi/2,H)$, and the transverse one, $\Delta\sigma(0,H)$. We self-consistently checked that the former terms are negligible. Within the two-dimensional approximation ($\gamma = 0$), the longitudinal magnetoconductance data in Fig.~\ref{MR} would directly yield the isotropic magnetoconductance. As the present compound is not highly anisotropic,\cite{rifi1994} the latter approximation is not strictly valid and it is necessary to determine the sample anisotropy. The anisotropic mass ratio is obtained in the following way: we first set $\Delta\sigma_\text{iso}\equiv 0$ and determine the $\gamma$ value which scales the angular data $\Delta\sigma(\theta,H)$ at $H = 3$~T and at $H = 6$~T, using Eq.~(\ref{scale}). Then, using the longitudinal and the transverse magnetoconductance data and Eqs.~(\ref{scalea}, \ref{scaleb}), the isotropic component is obtained as well as the orbital one. The scaling procedure is iterated taking into account the new orbital component until the $\gamma$ value is consistent with the set of angular- and field-dependent magnetoconductance data. As shown in Fig.~\ref{angular}, $\gamma^{-1}= 22 \pm 2$ scales the data nicely. We have used this anisotropy value to extract the isotropic and orbital component of the magnetoconductance for all doping states.

\section{Results and discussion}
\label{results}

\subsection{Positive isotropic magnetoconductance}

\begin{figure}
\includegraphics[width= \columnwidth]{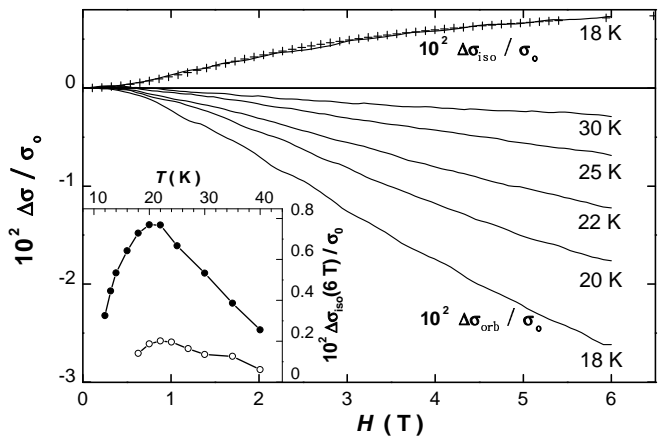}
\caption{Isotropic (positive) and orbital (negative) magnetoconductance vs $H$ for the most underdoped state (\textbf{j} in Fig.~\ref{MR}). The magnetoconductance is expressed in units of $\sigma_0 = e^2/16 \hbar s$, see equation~\ref{ALO}. The inset shows the isotropic magnetoconductance at 6~T vs $T$ (filled and open symbols are for the most (\textbf{j}) and second most underdoped state (\textbf{i}) in Fig.~\ref{MR} respectively). Crosses are a fit of the isotropic magnetoconductance using the `KMHZ' formula, with the Kondo temperature $T_K$ = 15~K and the crossover field $H_{cr}$ = 0.9~T.}\label{MRetat11}
\end{figure}

For the two most underdoped states (\textbf{i} and \textbf{j}), a positive magnetoconductance for field applied along the CuO$_2$ plane is found in a wide range of temperature (see the shaded area of Fig.~\ref{MR}). Using the procedure described above, it is found that this results from the existence of a positive, \textit{isotropic} magnetoconductance for these doping states (Fig.~\ref{MRetat11}). The origin of this positive contribution can be found neither in the superconducting fluctuations (since the only positive contribution to the field-dependent excess conductivity is the anomalous Maki-Thompson orbital one, which is anisotropic\cite{larkin2004}), nor in the two-dimensional weak localization (the latter was generally thought to explain the low-temperature upturn of the resistivity in the underdoped cuprates until the observation of a similar upturn in the transverse resistivity refuted this conventional picture\cite{ando1995}). 
The present results contrast with the ones obtained on a heavily underdoped, nonsuperconducting, Bi$_2$Sr$_2$CuO$_6$ single crystal in Ref.~\onlinecite{jing1991}. Indeed, based on the observation of a positive, \textit{anisotropic} magnetoresistance, the authors interpreted the low-temperature resistivity upturn as the occurrence of weak two-dimensional localization. Our results for the most underdoped states clearly indicate that the magnetoconductance is here a balance between a \textit{negative}, \textit{anisotropic} contribution (due to the superconducting fluctuations) and a \textit{positive}, \textit{isotropic} one, the former overcoming the latter as $T \rightarrow T_c$.

It is generally admitted that the isotropic magnetoconductance originates from a spin effect. While a positive contribution is in the case of out-of-plane conductivity naturally understood as the closing of the pseudogap by the magnetic field (at a characteristic field given by $g \mu_B H \approx k_B\,T^*$, where $T^*$ is the temperature for the opening of the pseudogap\cite{shibauchi2001}), the in-plane positive contribution cannot be accounted by the same effect, as the opening of the gap in the spin excitation spectrum results in an increase of the conductivity below $T^*$ and a negative magnetoconductance.\cite{ando1995} Such a behavior may be expected from the Kondo mechanism. This was first proposed to account for the isotropic positive magnetoconductance in both insulating and superconducting La$_{2-x}$Sr$_x$CuO$_{4+y}$\cite{preyer1991} (similar results were reported in the case of Bi$_2$Sr$_2$Ca$_{0.8}$Y$_{0.2}$Cu$_2$O$_{8+\delta}$\cite{thopart2000}). In the Kondo scenario, a crossover between a quadratic field dependence and a logarithmic one is expected when $k_B T \simeq g \mu_B H_\text{cr}$, where $g=2$ (from the generalized Hamann formula\cite{keiter1975} -- or `KMHZ' formula). However, in the references \onlinecite{preyer1991} and \onlinecite{thopart2000} the crossover field was found one order of magnitude smaller than this prediction. The same observation may be done in the present case, as we have $g \mu_B H_\text{cr}/k_B T \simeq 0.1$ (using $H_\text{cr}\simeq 0.9$~T at $T=18$~K, Fig.~\ref{MRetat11}). The same discrepancy is observed from the data in Ref.~\onlinecite{sekitani2003}, where the low-temperature resistivity upturn for heavily underdoped, non superconducting La$_{2-x}$Ce$_{x}$CuO$_4$ was fitted using the KMHZ formula : the KMHZ formula similarly fails to properly describe the negative magnetoresistance at the higher temperatures. 

An alternative explanation may be looked for within a spin-glass scenario, since, as pointed out in Ref.~\onlinecite{preyer1991}, spin glasses also exhibit an isotropic positive magnetoconductance.  There are indeed evidences for the occurrence of a spin glass system on the high-doping side of the antiferromagnetic phase, coexisting with the superconducting one (for a review, see Ref.~\onlinecite{kastner1998}). Metallic alloys exhibit a spin-glass state between the very dilute situation, showing the Kondo effect, and the concentrated one, for which the ordered magnetic state (ferromagnetic or antiferromagnetic) is found. The glass state allows to account for a crossover field (actually, the exchange field) as small as the one observed here, as $k_B T_g \simeq \mu_B H_\text{cr}$, using $T_g \simeq 1$~K, $T_g$ being the freezing temperature. However, in this model, for temperatures higher than $T_g$, the magnetoconductance should be quadratic up to a magnetic field much larger than $H_\text{cr}$:\cite{nigam1983,vegvar1991} this is clearly not observed here (Fig.~\ref{MRetat11}). 

Thus, \textit{for both the Kondo scenario and the spin-glass one, we are faced with a contradiction} concerning the energy scales inferred from the crossover field, and the much larger thermal energy for the temperature at which it is measured. Such a contradiction can be resolved only in the case of the existence of strong magnetic correlations. A remarkable feature is the occurrence of a maximum of the positive magnetoconductance at $T \simeq 20$~K (Fig.~\ref{MRetat11}, inset). A similar behavior is usually observed in ordered antiferromagnets close to $T_N$. Thus, this non monotonic behavior points towards the competition between a reduction of the spin scattering due to increased AF correlations and an increase due to local Kondo interactions as the temperature is lowered. 

Finally, it should be underlined that the negative magnetoresistance discussed above represents only a very small fraction of the low temperature resistivity upturn. The fit in Fig.~\ref{MRetat11} indicates that the magnetic field is very far from asymptotically exhausting this upturn, so the mechanisms discussed above do not allow us to conclude on the origin of the resistivity upturn.

\subsection{Orbital magnetoconductance}
\subsubsection{Dominant Aslamazov-Larkin contribution}

Two contributions to the orbital magnetoresistance are expected: the Aslamazov-Larkin one (ALO) and the Maki-Thompson one (MTO). Since the MTO contribution is less singular than the ALO one, it is expected to overcome the ALO contribution only well above $T_c$, as verified experimentally.\cite{hikami1988,matsuda1989} This situation should be accentuated here with respect to cuprates with a higher $T_c$, due to the combined effects of stronger disorder and lower temperature. In this regime, the characteristic field of the MTO contribution is $H_\phi = \hbar/(4\,D\,e\tau_\phi$), where $D$ is the diffusion coefficient and $\tau_\phi$ is the phase dephasing time.\cite{larkin1980} We evaluate this field within a crude cylindrical Fermi surface picture. Considering the optimally doped state (\textbf{e}, $T_c = 18.3$~K) and $T\simeq 2\,T_c$, using the carrier concentration at optimal doping $n\simeq 1.5\times 10^{27}$~m$^{-3}$ and the corresponding two-dimensional Fermi vector $k_F = 3.4\times 10^9$~m$^{-1}$ (this obtained from both the straightforward chemical valency computation and the thermoelectric power in Ref.~\onlinecite{konstan2003}, and about twice smaller than would be obtained from the maximum in the Hall resistance of this sample at about 100~K\cite{ando2000}) and the resistivity 
$\rho = 2.3\times 10^{-6}\;\Omega$m, we compute the metallic parameter $k_F\, l = 14$. The two-dimensional free electron model, for which $k_F\,l =  h s/\rho\,e^2$ -- where $s=12.3$~\AA{} is the conducting plane separation -- yields the same value. Taking for the hole effective mass $m^*=3\,m_e$,\cite{tsvetkov1997} we get the diffusion coefficient, $D=\hbar\,k_F\,l/(2\,m^*)=3\times 10^{-4}$~m$^2$/s. Being in the linear resistivity regime, we may evaluate the dephasing rate as the thermal transport scattering rate, \mbox{$\tau = 3\times 10^{-14}$~s}. From this, we obtain the dephasing field $H_\phi \simeq 20$~T. This is only a crude approximation and the Fermi surface complexity for cuprates must be taken into account to account for their transport properties.\cite{bok2004} Nevertheless, we consider that this is an indication that the contribution from the MTO process should be small in the range of field and temperature of interest. Furthermore, this correction, larger for larger temperature (the MTO contribution overcomes the ALO one as the temperature rises), should contribute to increase the magnetoresistance and cannot account for the curvature of the $H_{c2}^*(T)$ line discussed below.

\subsubsection{Crossover field}

The direct contribution of the thermodynamic fluctuations to the conductivity (the orbital Aslamazov-Larkin contribution -- ALO) is determined by the same coherence length as in the superconducting regime. Hence, a characteristic field $H_{c2}^*(T)$ can be determined as a mirror of the conventional upper critical field $H_{c2}(T)$. The formula describing the ALO contribution generally assume a linear dependence in $T$ for the characteristic field $H_{c2}^*$ (which corresponds to a square root divergence of the coherence length near $T_c$). We used the more general formula of the ALO contribution where the temperature dependence for $H_{c2}^*(T)$ is kept arbitrary:\cite{abrahams1971}
\begin{eqnarray}
\sigma_\text{ALO} &=& -2\, \sigma_0\; \epsilon^{-1}\Upsilon_\text{ALO}(H/H_{c2}^*(T)),
\nonumber\\
\Upsilon_\text{ALO}(x) &=& \{[\psi(1+1/(2x))-\psi(1/2+1/(2x))]/x-1\}/x,
\nonumber\\
\sigma_0 &=& e^2/16 \hbar s,
\label{ALO}
\end{eqnarray}
where $\sigma_0$ is proportional to the universal ALO excess conductance per square in a 2D superconductor $\sigma_\square$ ($\sigma_0=\sigma_\square/s$), 
$\psi$ is the digamma function, $\epsilon = \ln(T/T_c)$ and $H_{c2}^*(T)$ is the crossover field, symmetric with respect to $T_c$ of the upper critical field $H_{c2}(T)$. When $H \ll H_{c2}^*(T)$, the field dependence in Eq.~\ref{ALO} is quadratic and the magnetoconductance is, assuming $H_{c2}^*(T) = \epsilon\;H_{c2}^*(0)$:
\begin{equation}
\Delta\, \sigma_\text{ALO} \simeq - \sigma_0\,H^2/2 H_{c2}^*(0)\,\epsilon^3
\label{quadratic}
\end{equation}

\begin{figure}
\includegraphics[width= \columnwidth]{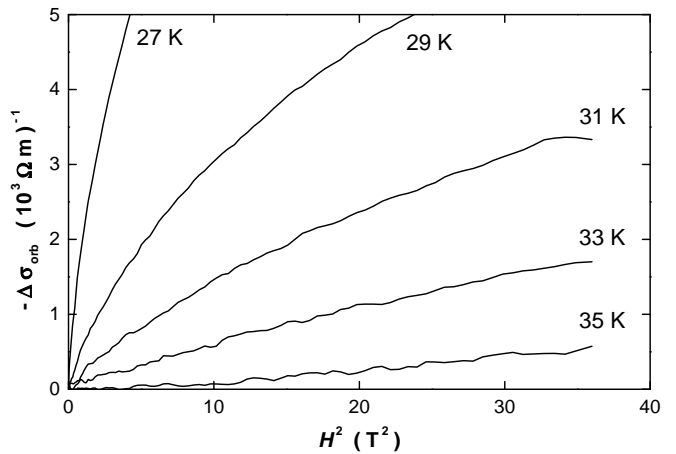}
\caption{Orbital magnetoconductance vs $H^2$ for the almost optimally doped state with $T_c$ = 18.3~K (state \textbf{e} in Fig.~\ref{MR}) showing quadratic field dependence at high temperature.}\label{MRetat6}
\end{figure}

\begin{figure}
\includegraphics[width= \columnwidth]{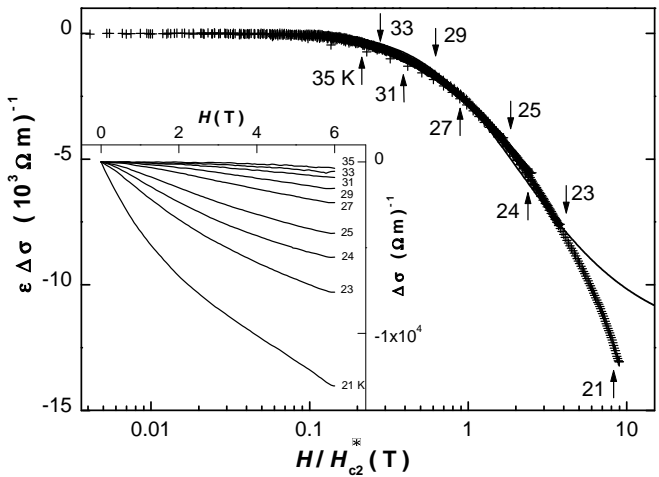}
\caption{Orbital magnetoconductance (state \textbf{e}) rescaled according to the ALO expression in Eq.~(\ref{ALO}). The deviation from Eq.~(\ref{ALO}) (line) is observed at $T\lesssim 21$~K and $H\gtrsim 3$~T. Arrows indicate the location of the $H = 6$~T point for several temperatures. The inset is the unscaled orbital magnetoconductance.}\label{fitALetat6}
\end{figure}

We do observe such a quadratic dependence at high temperature. A linear fit of the data in Fig.~\ref{MRetat6} ($T = 33$~K) straightforwardly provides the value $H_{c2}^*(0) = 37 \pm 4$~T for the almost optimally doped state (similarly, this procedure yields $H_{c2}^*(0)$ value for other doping states roughly proportional to $T_c$). This agrees with the determination of the upper critical field from resistivity data in the limit $T \rightarrow 0$ presented in Ref.~\onlinecite{osofsky1993}. However, it is observed that quadratic fits of the magnetoconductance for larger temperature yields larger values of the upper critical field. This means that $H_{c2}^*(T)$ is not a linear function of $\epsilon$ as assumed in Eq.~(\ref{quadratic}) and that it exhibits an upward curvature. This may be seen using the full expression in Eq.~(\ref{ALO}) to determine $H_{c2}^*(T)$. A convenient way to do this is to determine $H_{c2}^*(T)$ such that $\epsilon\;\Delta\,\sigma_\text{orb}$ vs $H/H_{c2}^*(T)$ obtained at different field and temperature values defines a single curve. As seen in Fig.~\ref{fitALetat6}, it is possible to reasonably scale the data along this scheme. The universal function in Eq.~(\ref{ALO}) roughly accounts for the scaled data. We have noticed, however, that for the strongly overdoped and underdoped states, the prefactor $\sigma_0$ in Eq.~(\ref{ALO}) -- when used as a fitting parameter -- is reduced by about 30\% with respect to the theoretical value, which may result from a spread of the doping level. Also, at low temperature ($\epsilon \lesssim 0.1$), the universal function obtained in this way systematically deviates from Eq.~(\ref{ALO}), which could be due to the finite width of the transition or to the occurence of critical fluctuations in the vicinity of $T_c$. The $H_{c2}^*$ values needed to scale the data define a strongly curved $H_{c2}^*(T)$ line, the curvature being stronger for lower doping (Fig.~\ref{Hc2vsT}). Given the reduced temperature and the transition temperature, the crossover field found in this way is also lower for the underdoped regime, which is a direct consequence of the excess magnetoconductance on the underdoped side (Fig.~\ref{excessMR}). This excess cannot be due to the onset of localization, which would result in a positive contribution. We have also checked, using the procedure described in Ref. \onlinecite{pomar1996}, that the effect of $T_c$ inhomogeneities is to depress the $H_{c2}(T)$ values obtained from our procedure only for temperatures such that $T-T_c \lesssim \Delta T_c$, where $\Delta T_c$ is the width of a Gaussian $T_c$ distribution. So, a curvature for $T$ larger than $2\,T_c$, as observed in the underdoped regime, cannot be accounted by a distribution of $T_c$. 

\subsubsection{Discussion}

\begin{figure}
\includegraphics[width= \columnwidth]{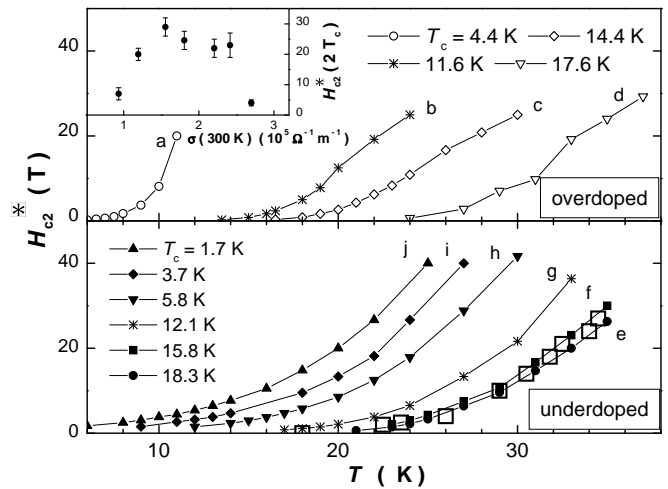}
\caption{Crossover field $H_{c2}^*$, obtained from the scaling of $\epsilon\,\Delta\,\sigma_\text{orb}$ \textit{above} $T_c$. Open squares are obtained from resistivity measurements \textit{below} $T_c$ in a magnetic field up to 20~T for a thin film with $T_c = 16.5$~K,\cite{rifi1996} following a mirror symmetry with respect to $T_c$ (this film was nearly optimally doped, in a state close to state \textbf{e} of our sample). Both procedures yield a similar curved crossover field. The inset shows that the crossover field obtained from fluctuations roughly scales as $T_c$.}\label{Hc2vsT}
\end{figure}

\begin{figure}
\includegraphics[width= \columnwidth]{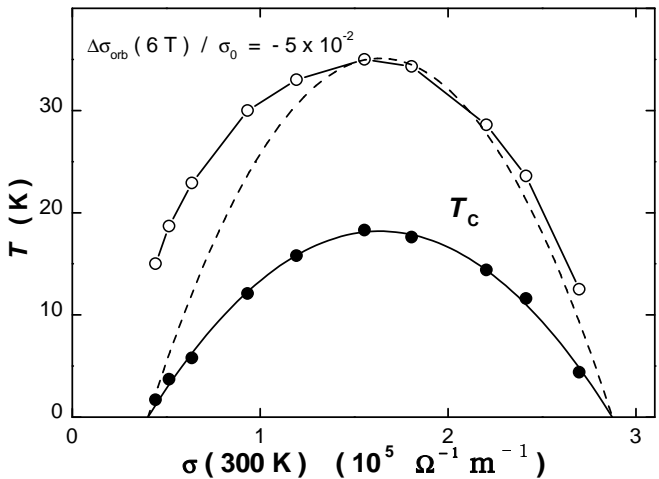}
\caption{Iso orbital magnetoconductance (open circles) and $T_c$ (full circles) in a temperature-$\sigma$(300~K) phase diagram for the different doping (see Fig.~\ref{MR} for comparison). The full and dotted lines are homothetic parabolas. The open circles represent the temperature where the orbital magnetoconductance at our maximal field reaches a given value, expressed as a ratio of $\sigma_0$ (see Eq.~\ref{ALO}). Changing this ratio ($-5\times10^{-2}$ here) does not alter the shape of the dome. An excess of negative orbital magnetoconductance can be seen on the underdoped side.}\label{excessMR}
\end{figure}

The $H_{c2}^*(T)$ found in this way is strikingly similar to the one inferred from the onset of the resistive transition. We show for comparison in Fig.~\ref{Hc2vsT} the $H_{c2}(T)$ values obtained  \textit{below} $T_c$ on a similar sample in a magnetic field up to 20~T,\cite{rifi1994,rifi1996} using the determination of Ref.~\onlinecite{osofsky1993}. Our data obtained \textit{above} the zero field transition temperature confirm this curvature. It appears also that the transition is less robust to the magnetic field on the underdoped regime than it is on the overdoped one: the curvature for $H_{c2}^*(T)$ appears to be stronger for the underdoped states (this is best evidenced in Fig.~\ref{Hc2vsT}, by comparing the overdoped state \textbf{b}, with $T_c = 11.6$~K, to the corresponding underdoped state \textbf{g}, with $T_c = 12.1$~K). 

In a general manner, the observation of symmetric lines in the superconducting and the fluctuation regimes points toward the existence of a conventional correlation length similar to that obtained from the mean field theory of second-order phase transitions (defined as $\xi^2=\hbar /( 2\,e\,H_{c2})$), which is probed by the magnetic field. We note that Nernst-effect measurements may also be used to define symmetric crossover lines. Such measurements locate the $H_{c2}(T)$ line obtained from resistivity measurements as a crossover between the melting line and the ridge line joining points of maxima of the Nernst signal vs $H$.\cite{wang2005} A ridge line may also be found in the fluctuation regime above $T_c$ from the data in Fig.~13 of Ref.~\onlinecite{wang2005}: this line -- connecting the points where the contour line of the Nernst signal shows a vertical tangent -- is found roughly symmetric with respect to $T_c$ and may indicate the existence of a similar crossover line in the fluctuation regime. 

We believe our observation first rules out a conventional flux-flow mechanism in a vortex line liquid as the origin of the anomalous $H_{c2}(T)$ curvature. Indeed, this should occur in the vicinity of the vortex melting line transition (in the ($H,T$) plane), which is first order and thus does not have a dual line in the fluctuation regime. However, above this line, it was proposed from numerical simulations that there exists a second line which may be viewed as an extension of the zero field vortex loop unbinding transition to finite magnetic field.\cite{nguyen1999} Such a vortex unbinding is invoked to account for the anomalous Nernst signal. According to Ref.~\onlinecite{nguyen1999}, this line may represent a true thermodynamic transition, while the mean field $H_{c2}(T)$ line would be reduced to a crossover line. Now, does this transition define a correlation length that could also be probed by the magnetic field in the regime of fluctuating Cooper pairs? By analogy with the low-temperature mechanism, this requires that fluctuating vortex loops can thread the superconducting fluctuating domains which has, to our knowledge, never been considered.

Description of high-$T_c$ superconductivity by means of granular models is appealing, as there is some evidence that the granular nature of the Bi-based compounds is stronger in the underdoped regime,\cite{davis2005,mashima2006} which would be in agreement with the marked curvature of the $H_{c2}^*(T)$ line in this regime. Also, such a granularity may account for the observation of an anomalous effect of the transport current on the superconducting fluctuations in Bi-2201.\cite{sfar2005} Concerning the superconducting cluster model,\cite{spivak1995,geshkenbein1998} the proposed mechanism -- the magnetic-field induced decoupling of large superconducting islands, connected through a normal-metal proximity effect -- implies that the resistive transition line is governed by the temperature for coherence of the normal-metal areas, which is much smaller than the phase-ordering temperature. So, this model cannot account for the existence of a symmetric line $H_{c2}^*(T)$ in the fluctuating regime. A similar objection may be made against the phase fluctuation model in Ref.~\onlinecite{emery1995b}: when the mean-field temperature at which pairing sets in is well above the transition temperature for phase coherence, the analogy of this model with one of a Josephson-junctions array allows to conclude that no symmetric transition line is expected with respect to the phase-coherence transition temperature. The model in Ref.~\onlinecite{ovchinnikov1996} for inhomogeneities of the order of the coherence length does not have such an inconvenient, but corrections to the conventional upper critical field are small and leave this quantity unchanged near $T_c$. So it is difficult to account for the reduced slope of the $H_{c2}^*(T)$ line close to $T_c$.

Concerning the magnetic impurities mechanisms,\cite{ovchinnikov1996,ovchinnikov1999} the pair-breaking effect must be specific to the underdoped regime. As it is essential also in this model, the spin-flip scattering time should increase as the temperature decreases: our observation of a positive isotropic magnetoconductance in the strongly underdoped regime and a possible magnetic ordering at low temperature goes along these lines. However, the fact that this mechanism rests on some low magnetic-ordering temperature breaks the symmetry with respect to $T_c$, and the $H_{c2}(T)$ curvature should actually be reversed in the fluctuation regime.

Finally, the increased curvature in the underdoped regime appears to be consistent with both the Bose-Einstein condensation and the Boson fermion models. As can be seen from the above considerations, it is essential that, in these models, the upper critical field is determined only by the bulk transition temperature and that no other temperature scale (such as magnetic ordering in Ref.~\onlinecite{ovchinnikov1996}, or coherence of a metallic grain in Ref.~\onlinecite{geshkenbein1998}) is found. In the case of the models with preformed Bosons,  there is an energy scale corresponding to their formation; but this energy scale does not appear in the determination of the upper critical field, making these models compatible with the present observation.

\section{Conclusion}

We have investigated the magnetoresistance for a Bi$_2$Sr$_2$CuO$_{6+\delta}$ thin film, from highly overdoped to highly underdoped states above the zero-field resistive superconducting transition temperature. The isotropic magnetoresistance is found negative for the lower (but still superconducting) doping states, which we tentatively interpret as a contribution from strongly correlated magnetic scatters. The orbital positive magnetoresistance is fitted by the Aslamazov-Larkin theory. This yields an anomalous critical-field temperature dependence, which agrees with previous resistive measurements below $T_c$. This points toward the existence of a similar correlation length above and below $T_c$, as expected for a continuous transition and rules out models where the resistive transition is determined by the onset of phase coherence, as is the case for the flux lattice melting mechanism or the phase decoupling one.

\begin{acknowledgments}

We acknowledge the support of CMCU to Project No. 04/G1307.
\end{acknowledgments}

\end{document}